\documentclass[a4paper]{jpconf}
\usepackage{graphicx}

\def\be{\begin{equation}}
\def\ee{\end{equation}}
\def\bea{\begin{eqnarray}}
\def\eea{\end{eqnarray}}

\def\f{\frac}
\def\s{\sqrt}

\def\l{\left}
\def\r{\right}

\def\k{\kappa}

\def\d{{\rm d}}
\def\7{$\;$}
\def\p{\partial}

\begin{document}
\title{Palatini NMDC gravity: cosmological scalar field phase portraits in exponential potential}

\author{Candrasyah Muhammad, Somphoach Saichaemchan \\ and Burin Gumjudpai}
\address{The Institute for Fundamental Study ``The Tah Poe Academia Institute" \\ Naresuan University, Phitsanulok 65000, Thailand}
\ead{candrasyahm59@email.nu.ac.th , somphoachs58@email.nu.ac.th}
\ead{buring@nu.ac.th}
\begin{abstract}
We consider cosmological scalar field evolving under exponential potential of the Non-minimal Derivative Coupling (NMDC) gravity model in Palatini formalism.  Slow-roll regime is assumed.   GR and metric formalism NMDC cases are compared in this study. Phase portraits show that Palatini NMDC effect restricts acceleration phase into smaller region in the phase space.  NMDC effect of the Palatini case
suppresses expansion rate than that of the GR while the metric NMDC
enhances rate of expansion of the GR case.
\end{abstract}

\section{Introduction}
Present acceleration of the universe is confirmed by many astrophysical observations, such as Supernova type Ia (SN Ia) \cite{Perlmutter:1997zf, Riess:1998cb, Astier:2005qq}, Cosmic Microwave Background (CMB) Anisotropies \cite{Ade:2015lrj}, large-scale structure surveys, and X-ray luminosities from galaxy clusters \cite{Rapetti:2005a}. The dark energy is believed to be responsible for the acceleration. Dark energy could be in the form of dynamical fields, therefore many scalar field models are proposed in order to explain the present accelerated expansion (see e.g. \cite{Copeland:2006a} and references therein) as well as inflation in the early universe \cite{Guth, RP1988}.
Modifications of the gravitational theory can be performed by changing geometrical sector in the Einstein-Hilbert action of which the acceleration can be attained in many ways at both early and late times \cite{Carroll2004, SCVF}. Coupling between scalar sector and geometry can be found in classes of scalar-tensor theories \cite{maeda}. These contain non-minimal coupling (NMC) term between the Ricci scalar and a scalar field (see e.g. \cite{Amendola1993}).  The NMC model can be  extended to coupling of the Einstein tensor to derivative of the scalar field dubbed a non-minimal derivative coupling (NMDC) theory \cite{Capozziello:1999xt, Granda:2010hb, Sushkov:2009, Saridakis:2010mf, Germani:2010gm, Sushkov:2012, Tsujikawa:2012mk, Sadjadi:2010bz, Gumjudpai:2015vio, Gumjudpai:2016frh}. The NMC and NMDC are found as a subclass of Horndeski’s theory \cite{Horn} which is the most generalized case of the action with at most second-order derivative of the metric and of the scalar field.
Metric NMDC theory allows inflation by enhanching friction even in steep potentials with theoretically natural models parameters and helps supressing the tensor-to-scalar ratio \cite{Tsujikawa:2012mk}. A version of NMDC theory as proposed by \cite{Sushkov:2012} was investigated in metric formalism which regards the metric as a dynamical field. Alternative ideas lies on symmetrical properties of the manifold space. Riemannian manifold possesses isometry property and the metrical field is consistent with it. A space that possesses the diffeomorphism symmetry, allows connection and the metric to be separated dynamical fields so that these fields are varied independently. This approach is called Palatini formalism which gives different equations of motion from the metric approach unless considering GR \cite{Olmo:2011uz}.
The Palatini NMDC action is given by
\begin{equation}
\tilde{S}[g, \Gamma, \phi] = \frac{M_{\rm P}^2 c^4}{2} \int \d^4 x \s{-g} \l[\tilde{R}-\l(\varepsilon g_{\mu\nu}+\k\tilde{G}_{\mu\nu}\r)\phi^{,\mu}\phi^{,\nu}   - 2V \r]\,, \label{PalatiniAct}
\end{equation}
where $c \equiv 1$, $\tilde{G}_{\mu\nu}(\Gamma) = \tilde{R}_{\mu\nu}(\Gamma)-\f{1}{2} g_{\mu\nu} \tilde{R}(\Gamma)$; $\tilde{R}(\Gamma) = g^{\mu\nu}\tilde{R}_{\mu\nu}(\Gamma)$ and $\tilde{R}_{\mu\nu}(\Gamma) = R^{\lambda}_{\mu\lambda\nu}(\Gamma) = \p_{\lambda}\Gamma^{\lambda}_{\mu\nu}-\p_{\nu}\Gamma^{\lambda}_{\mu\lambda} + \Gamma^{\lambda}_{\sigma\lambda}\Gamma^{\sigma}_{\mu\nu} - \Gamma^{\lambda}_{\sigma\nu}\Gamma^{\sigma}_{\mu\lambda}$ with detailed as referred in \cite{Gumjudpai:2016ioy}\footnote{A different version of Palatini NMDC was investigated before in \cite{palaNMDC}.}. The  matter fields are negligible here.
Previous works \cite{Gumjudpai:2016ioy, Saichaemchan:2017psl} showed that positive $\k$ results in superluminal graviton speed. Albeit the $V \propto \phi^4$ potential with $\k > 0$ could pass the CMB constraint \cite{Ade:2015lrj}, it is with only by large amount of fined-tuning. Moreover, the negative $\k$ case with $V \propto \phi^2$ fails the CMB constraint. Hence both $V \propto \phi^2 $ and $ V \propto \phi^4$ are not likely to be viable. With this reason, we consider $\k < 0$ and use $V = V_0 e^{-\lambda \phi / M_{\rm P}}$ instead of the power-law (chaotic) potential.

\section{Equations of motion in the slow-roll regime}
Giving the scenario of the NMDC gravities, it is interesting to view comparative graphical presentations of the dynamics. In doing such, we consider sets of the autonomous system in GR, metric NMDC and Palatini NMDC gravities in slow-roll regime so that the slow-roll approximations  $0 \ll |\phi^2 | \ll 1; \ |\ddot{H}/H| \ll |\dot{H}| \ll |H^{2}|; \ |4\dot{H}\kappa| \ll 1,$ can be applied \cite{Gumjudpai:2016ioy}.
\subsection{Metric NMDC Gravity}
Considering scalar field  as the only species in inflationary epoch, $\rho_{\rm tot} \equiv \rho_{\phi} = \varepsilon\dot{\phi}^2/2 + V ; \ p_{\rm tot} \equiv p_{\phi} = \varepsilon\dot{\phi}^2/2 - V $. Friedmann  and the Klein-Gordon equations are \cite{Gumjudpai:2015vio}
\be
H^2 = \f{1}{3 M_{\rm P}^2}\l[  \f{\dot{\phi}^2}{2}\l(\varepsilon  - 9 \kappa H^2 \r)  + V(\phi)  \r]\,, \label{FRMetric}
\ee
\be
\ddot{\phi}\l(\varepsilon - 3 \kappa H^2  \r) +
3 H \dot{\phi}\l(   \varepsilon -  3 \kappa H^2    \r)  + V_{\phi} \simeq   0\,.  \label{KGMetric}
\ee
One can set an autonomous system:
\bea
\dot{\phi}  =     \psi \,,\;\;\;
\dot{\psi}    \simeq      \f{-V_{\phi}  - 3 H \psi (\varepsilon - 3 \k H^2)  }{\varepsilon - 3 \k H^2} \,,\;\;\;
\dot{H}      \simeq     \f{V_{\phi} \psi }{6 M_{\rm_P}^2 H}\,. \label{autoMetric}
\eea

\subsection{Palatini NMDC Gravity}
From the Palatini action (\ref{PalatiniAct}), the Friedmann and the Klein-Gordon equations are given by
\be
  H^2  \simeq   \f{1}{3 M_{\rm P}^2}\l[  \f{1}{2}\varepsilon\dot{\phi}^2  + V(\phi)   \r]\,,   \label{FRPala}
\ee
\be
\ddot{\phi}\l[ \varepsilon   -  {\f{15}{2} \kappa}H^2 \r]  +  3H \dot{\phi} \varepsilon    + V_{\phi}  \;  \simeq \;  0\,.  \label{KGPala}
\ee
An autonomous system can be derived as follows :
\bea
\dot{\phi}  =      \psi\,,\;\;\;
\dot{\psi}   \simeq     \f{-V_{\phi}  - 3 H \psi \varepsilon   }{\varepsilon - (15/2) \k H^2}\,,\;\;\;
\dot{H}      \simeq    \f{V_{\phi} \psi  }{6 M_{\rm_P}^2 H}\,. \label{autoPala}
\eea

\section{Acceleration condition for exponential potential}
Exponential potential has similar motivation as the NMDC models, i.e. it is motivated by low-energy effective theories of quantum gravities \cite{Copeland:1997et}. Therefore in this study, we consider exponential potential,
$ V = V_{0}\exp(-\lambda\phi/M_{\rm P})\,.
$
Acceleration condition is obtained by taking $\ddot{a}/a \equiv \dot{H} + H^2 > 0$.

\subsection{Metric NMDC Gravity}
In the metric case, we use approximation $H \approx \s{V}/\l(\s{3}M_{\rm P}\r)$ and $|\k| \ll |M_{\rm P}^2/V|$ to obtain the acceleration condition,
\bea
		\dot{\phi}^2 \; < \; V_{0}\exp\left(-\f{\lambda\phi}{M_{\rm P}}\right)\l\{1-\frac{\kappa \varepsilon\lambda^2 V_{0} \exp\left(-\f{\lambda\phi}{M_{\rm P}}\right)}{M_{\rm P}^2\left[\varepsilon - \frac{\kappa}{M_{\rm P}^2}V_{0}\exp\left(-\f{\lambda\phi}{M_{\rm P}}\right)\right]}\r\}\l[\varepsilon-3\kappa\frac{V_{0}\exp\left(-\f{\lambda\phi}{M_{\rm P}}\right)}{M_{\rm P}^2}\r]^{-1}\,. \label{accelmetric}
\eea


\subsection{Palatini NMDC Gravity}
Considering binomial approximation for $|\kappa| \ll |M_{\rm P}^2 /V|$ and late time $\dot{\phi} \approx -M_{\rm P}V_{\phi}/\sqrt{3}\epsilon\sqrt{V}$, acceleration condition is given by
\begin{equation}
\varepsilon\dot{\phi}^2 \;< \;  V_{0}\exp\left(-\f{\lambda\phi}{M_{\rm P}}\right)\left\{ 1 + \frac{\kappa\lambda^4}{\varepsilon M_{\rm P}^2}V_{0}\exp\left(-\f{\lambda\phi}{M_{\rm P}}\right)\left[ 1+\frac{15\varepsilon\kappa}{8M_{P}^2}V_{0}\exp\left(-\f{\lambda\phi}{M_{\rm P}}\right)\right]\right\}\,. \label{accelPala}
\end{equation}

\section{Results}
\begin{figure}[h]
\centering
\includegraphics[width=41pc]{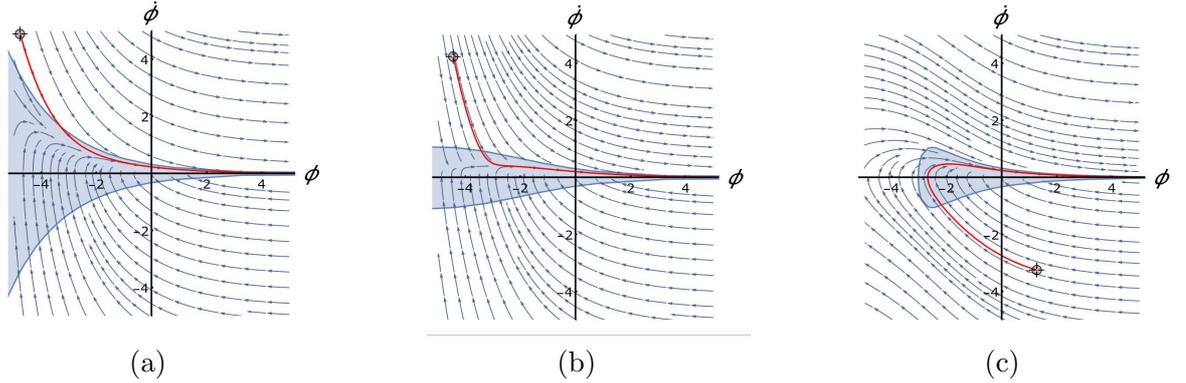}\vspace{0.5pc}
\put(-420,0){(a)}
\put(-260,0){(b)}
\put(-100,0){(c)}
\caption{\label{fig1}Phase portrait for an NMDC scalar field under exponential potential, with parameter $V_{0} = 1; \ \kappa = -0.5; \varepsilon = 1; \ M_{\rm P} = 1.0; \ \lambda = 1.0 $ for three cases: (a) GR  (b) metric NMDC and (c) Palatini NMDC cases; red curves are trajectories from some chosen initial conditions.
}
\end{figure}

\begin{figure}[h]
\centering
\includegraphics[width=26pc]{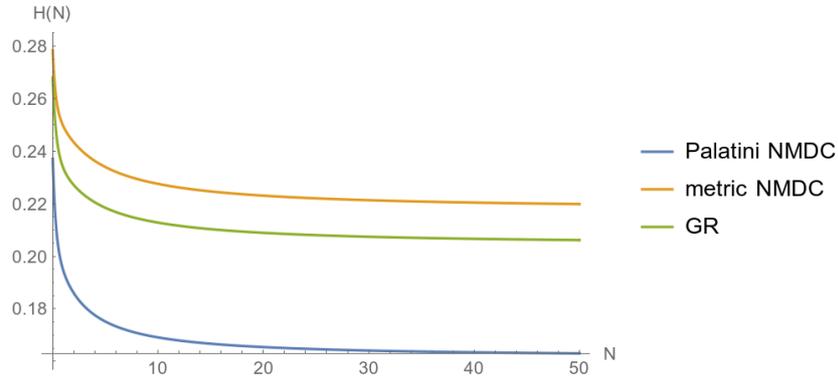}\vspace{0.5pc}
\caption{\label{fig2} $H$ plotted over e-folding number $N$ in GR, metric NMDC and Palatini NMDC cases.}
\end{figure}

We plot the phase portraits of the 3-dimensional autonomous systems both of the metric (Eq.(\ref{autoMetric})) and the Palatini (Eq. (\ref{autoPala})) cases. The GR case is portraited by taking $\kappa = 0$. Fig. \ref{fig1} shows the phase portrait for the three cases and the acceleration region (as described by Eqs. (\ref{accelmetric}) and (\ref{accelPala})) are presented. Fig. \ref{fig2} shows evolution of $H$ as e-folding number $N$ grows in the three cases.

\section{Conclusion}
We present phase portraits of non-minimal derivative coupling scalar field under exponential potential, $V = V_{0}\exp(-\lambda\phi/M_{\rm P})$ in metric and
Palatini formalism. Previous works \cite{Gumjudpai:2016ioy, Saichaemchan:2017psl} showed difficulty of the theory to be viable with chaotic inflation potentials, motivating consideration of exponential potential with the negative $\kappa$ case of the theory.
In the slow-rolling regime, Acceleration
conditions are derived in three cases. As seen in Fig. \ref{fig1}, there is only one attractor in all cases. However, Palatini NMDC gives smaller acceleration region and restricts the attractor into that smaller region. It is seen from Fig. \ref{fig2} that Palatini NMDC effect suppresses expansion rate than that of the GR case while the metric NMDC effect enhances the expansion rate from that of the GR case.

\section*{Acknowledgements} This work is supported by Naresuan University Research Grant-R2557C121.

\end{document}